\documentclass{iau-JDSS}

\title[LOFAR] 
{A very brief description of LOFAR -- the Low Frequency Array}

\author[Falcke et al.]   
{H.~Falcke$^{1,2,\mbox{\thanks{Visiting Miller Professor, Astronomy Department, University of California at Berkeley}}}$,
M.P.~van~Haarlem$^1$, A.G.~de~Bruyn$^{1,3}$, R.~Braun$^1$, H.J.A.~R\"ottgering$^4$, B.~Stappers$^{1,5}$, 
W.H.W.M.~Boland$^4$, H.R.~Butcher$^1$,  E.J.~de~Geus$^1$, L.~Koopmans$^3$, R.~Fender$^{5,6}$, 
J.~Kuijpers$^2$, G.K.~Miley$^4$, R.T.~Schilizzi$^{7,4}$, C.~Vogt$^1$, R.A.M.J.~Wijers$^5$,  M.~Wise$^{5}$, 
W.N.~Brouw$^3$, J.P.~Hamaker$^1$, J.E.~Noordam$^1$, T.~Oosterloo$^1$,
L.~B\"ahren$^{1,2}$, M.A.~Brentjens$^{1,3}$, S.J.~Wijnholds$^1$, 
J.D.~Bregman$^1$,  W.A.van Cappellen$^1$, A.W.~Gunst$^1$, G.W. Kant$^1$, J.~Reitsma$^1$, K.~van~der~Schaaf$^1$, C.M.~de~Vos$^1$
}

\affiliation{$^1$ASTRON, Postbus 2, 7990 AA Dwingeloo, NL
\\[\affilskip]
$^2$Department of Astronomy, Radboud University, Postbus 9010, 6500 GL Nijmegen, NL\\[\affilskip]
$^3$Kapteyn Astronomical Institute, University~of~Groningen, PO Box 800, 9700 AV Groningen, NL\\[\affilskip] 
$^4$Leiden Observatory, Leiden University, PO Box 9513, 2300 RA Leiden, NL\\[\affilskip] 
$^5$Astronomical Institute "Anton Pannekoek," University of Amsterdam, Kruislaan 403, 1098 SJ Amsterdam, NL\\[\affilskip] 
$^6$School of Physics \& Astronomy, University of Southampton, Southampton, SO17 1BJ, UK
$^7$International SKA Project Office, c/- ASTRON, Postbus 2, 7990 AA Dwingeloo, NL
}

\pubyear{2006}
\volume{Volume 14}  
\pagerange{119--126}
\date{?? and in revised form ??}
\jname{Highlights of Astronomy, Volume 14}
\editors{K.A. van der Hucht, ed.}
\begin{document}

\maketitle

\begin{abstract}
LOFAR (Low Frequency Array) is an innovative radio telescope optimized
for the frequency range 30-240 MHz. The telescope is realized as a
phased aperture array without any moving parts. Digital beam forming
allows the telescope to point to any part of the sky within a
second. Transient buffering makes retrospective imaging of
explosive short-term events possible.

The scientific focus of \mbox{LOFAR} will initially be on four key
science projects (KSPs): 1) detection of the formation of the very
first stars and galaxies in the universe during the so-called epoch of
reionization by measuring the power spectrum of the neutral hydrogen
21-cm line (\nocite{ShaverWindhorstMadau1999}Shaver et al. 1999) on
the $\sim5^\prime$ scale; 2) low-frequency surveys of the sky with of
order $10^8$ expected new sources; 3) all-sky monitoring and detection
of transient radio sources such as gamma-ray bursts, x-ray binaries,
and exo-planets (Farrell et al. 2004)\nocite{FarrellLazioZarka2004};
and 4) radio detection of ultra-high energy cosmic rays and neutrinos
(\nocite{FalckeGorham2003}Falcke \& Gorham 2003) allowing for the
first time access to particles beyond $10^{21}$ eV (Scholten et
al. 2006\nocite{ScholtenBacelarBraun2006}). Apart from the KSPs open
access for smaller projects is also planned. Here we give a brief
description of the telescope.
\end{abstract}

\firstsection 
\section{LOFAR - how it works}
 In its first phase \mbox{LOFAR} will consist of 77 {\it stations}
 distributed within a ring of $\sim100$ km diameter. 32 stations will
 be clustered in a central core of $\sim$2 km diameter located in the
 North-Eastern part of the Netherlands near the village of Exloo. Each
 station has two antenna systems: the Low-Band and High-Band Antennas
 (LBA, HBA). The LBA system operates primarily in the frequency range
 30-80 MHz with a switch to observe over a 10-80 MHz band as well. The
 HBA is optimized for the range 110-240 MHz with a possibility to
 observe up to 270 MHz with lower sensitivity. The LBA field is 60 m
 in diameter and contains 96 inverted-V crossed dipoles oriented NE-SW
 and SE-NW (i.e. dual polarization) in a randomized distribution with
 a slight exponential fall-off in density with radius. The HBA field
 consists of 96 tiles distributed in an as yet undetermined manner
 over roughly 50m. Each tile consists of a 4$\times$4 array of
 bowtie-shaped crossed dipoles with an analog 5-bit beam former using
 true time delays. Radio waves are sampled with a 12-bit A/D-converter
 -- to be able to cope with expected interference levels -- operating
 at either 160 or 200 MHz in the first, second or third Nyquist zone
 (i.e., 0-100, 100-200, or 200-300 MHz band respectively for 200 MHz
 sampling). The data from the receptors is filtered in $512\times195$
 kHz sub-bands (156 kHz subbands for 160 MHz sampling) of which a
 total of 32 MHz bandwidth (164 channels) can be used at any
 time. Data from each receptor can be buffered in a transient buffer
 board (TBB) for as long as $\sim10$min$/(\Delta \nu/196$kHz). Subbands
 from all antennas are combined on station-level in a digital
 beamformer allowing 8 independently steerable beams which are sent to
 the central processor via a glas fibre link that handles 0.7 Tbit/s
 data. The beams from all stations are further filtered into 1 kHz
 channels, cross-correlated and integrated on typical timescales of
 1-10 seconds. The integrated visibilities are then calibrated on 10
 second intervals to remove the effects of the ionosphere and images
 are produced. Channels with disturbing radio frequency interference
 (RFI) are dropped. For the correlation we use four of the six racks
 of an IBM Blue Gene/L machine in Groningen with a total of
 $\sim$12000 processors. We expect a typical input rate of
 $\sim0.5$Tbit/s into the computer. Unix clusters are used as input
 and output nodes for pre- and postprocessing.
 
 The expected 3$\sigma$ point source sensitivities of \mbox{LOFAR} for one hour
 integration over 4 MHz bandwidth dual-polarization are 2 mJy, 1.3 mJy, 70 $\mu$Jy, and
 60 $\mu$Jy at 30, 75, 120, \& 200 MHz respectively. The resolution
 will be 25$^{\prime\prime}$, 10$^{\prime\prime}$, 6$^{\prime\prime}$,
 and 3.5$^{\prime\prime}$ for the same frequencies. The field of view
 is 3$^\circ$ at 150 MHz (HBA) and 7.5$^\circ$ at 50 MHz (LBA).

 The project is currently in discussions with consortia in Germany,
 UK, France, Italy, and Sweden to expand the baseline and increase the
 resolution by up to a factor of ten.

National funding of \mbox{LOFAR} has been obtained at a level of $\sim$75
MEuro plus various in-kind contributions. Construction of the first
station was completed in September 2006. Further stations are expected
to be rolled out in the course of 2007, so that commissioning and
start of operation is foreseen for the year 2008.

\section{Outlook and Conclusions}\label{sec:concl}
With its new concept (Bregman 2000\nocite{Bregman2000}) of a
broad-band aperture array and digital beamforming \mbox{LOFAR} is
expected to pave the way for a new generation of telescopes and to be
an important pathfinder for the Square Kilometre Array. \mbox{LOFAR}
will improve the resolution and sensitivity of previous telescopes for
continuum observations by roughly two orders of magnitude over a wide
frequency range. It will also provide instantaneous access to a large
fraction of the sky at low frequencies at once, making serious and
regular all-sky radio monitoring possible for the first time. With
these unusual properties \mbox{LOFAR} promises a wealth of new
discoveries.

\end{document}